\documentclass[english,10pt,aps,pra,twocolumn,superscriptaddress,floatfix,sort]{revtex4-1}

\usepackage{amsmath}
\usepackage{amssymb}
\usepackage{amsfonts}
\usepackage{bbm}
\usepackage{epsfig}
\usepackage{grffile}
\usepackage{babel}

\usepackage{color}
\definecolor{grey}{rgb}{.5,.5,.5}
\definecolor{dblue}{rgb}{0,0,.5}

\usepackage[colorlinks=true,citecolor=dblue,linkcolor=dblue,urlcolor=dblue]{hyperref}
\usepackage[all]{hypcap}

\newcommand{\id}{\mathbbm{1}}
\newcommand{\bra}{\langle}
\newcommand{\ket}{\rangle}

\renewcommand{\vec}[1]{{\boldsymbol{#1}}}

\newcommand{\norm} [1]{\left\Vert #1\right\Vert}

\newcommand{\ua}{\uparrow}

\newcommand{\vs}{\vec{\sigma}}
\newcommand{\bs}{\bar{\sigma}}
\newcommand{\vn}{\vec{n}}
\newcommand{\trunc}{\text{trunc}}
\newcommand{\SVD}{\text{SVD}}

\newcommand{\Section}[1]{\emph{#1.}~--}
\usepackage{xr}

\newcommand{\duke} {\fontfamily{ptm}\selectfont Department of Physics, Duke University, Durham, NC 27708, USA}
\newcommand{\santafe} {\fontfamily{ptm}\selectfont Santa Fe Institute, 1399 Hyde Park Road, Santa Fe, NM 87501, USA}
\newcommand{\lptms} {\fontfamily{ptm}\selectfont Laboratoire de Physique Th\'{e}orique et Mod\`{e}les Statistiques, Universit\'{e} Paris-Sud, CNRS UMR 8626, 91405 Orsay Cedex, France}

\newcommand{\Title} {A matrix product algorithm for stochastic dynamics on networks, applied to non-equilibrium Glauber dynamics}
\newcommand{\Authors}
{
\author{Thomas Barthel}
\affiliation{\duke}
\affiliation{\lptms}
\author{Caterina De~Bacco}
\affiliation{\santafe}
\affiliation{\lptms}
\author{Silvio Franz}
\affiliation{\lptms}
}
\newcommand{\Date} {November 5, 2017}

\begin{document}

\title{\Title}
\Authors
\date{\Date}

\begin{abstract}
We introduce and apply a novel efficient method for the precise simulation of stochastic dynamical processes on locally tree-like graphs. Networks with cycles are treated in the framework of the cavity method. Such models correspond, for example, to spin-glass systems, Boolean networks, neural networks, or other technological, biological, and social networks. Building upon ideas from quantum many-body theory, the new approach is based on a matrix product approximation of the so-called edge messages -- conditional probabilities of vertex variable trajectories. Computation costs and accuracy can be tuned by controlling the matrix dimensions of the matrix product edge messages (MPEM) in truncations. In contrast to Monte Carlo simulations, the algorithm has a better error scaling and works for both, single instances as well as the thermodynamic limit. We employ it to examine prototypical non-equilibrium Glauber dynamics in the kinetic Ising model. Because of the absence of cancellation effects, observables with small expectation values can be evaluated accurately, allowing for the study of decay processes and temporal correlations.
\end{abstract}


\maketitle

\Section{Introduction}
In recent years, we have seen increased efforts by statistical physicists to tackle stochastic dynamical processes in networks in order to study various phenomena \cite{Barrat2008,Newman2010} such as ordering processes, the spreading of epidemics and opinions, synchronization, collective behavior in social networks, stability under perturbations, or avalanche dynamics.

A drastic simplification can be achieved when short cycles in the network, defined by interaction terms, are very rare. This is the case for locally tree-like graphs such as random regular graphs, Erd\H{o}s-R\'{e}ny graphs, and Gilbert graphs. For such random graphs with $N$ vertices, almost all cycles have length $\gtrsim\log N$ such that their effect is negligible in the thermodynamic limit \cite{Mezard2009}. For static problems, this has been exploited in the so-called cavity method \cite{Mezard2001-20}, where conditional nearest-neighbor probabilities are computed iteratively within the Bethe-Peierls approximation. The method was very successfully applied to study, for example, equilibrium properties of spin glasses \cite{Mezard2001-20}, computationally hard satisfiability problems \cite{Mezard2002-297,Mezard2002-66}, and random matrix ensembles \cite{Rogers2008-78}.

This big success has motivated the generalization of the cavity method to dynamical problems \cite{Neri2009-08,Karrer2010-82}, which is known as the dynamic cavity method or dynamic belief propagation. Unfortunately, the number of possible trajectories and, hence, the computational complexity increase exponentially in time. Applications have hence been restricted to either very short times \cite{Neri2009-08,Kanoria2011-21}, oriented graphs \cite{Neri2009-08}, or unidirectional dynamics with local absorbing states \cite{Karrer2010-82,Lokhov2015-91,Altarelli2014-112,Shrestha2014-89}. In the latter case, one can exploit that vertex trajectories can be parametrized by a few switching times. The problem is hence to find good approximations to the exact solution of the dynamic cavity equations with polynomial computations costs.
Simple approaches are to neglect temporal correlations completely as in the one-step method \cite{Neri2009-08,Aurell2011-04,Aurell2012-85,Zhang2012-148} or to retain only some $\Delta t=1$ correlations as in the one-step Markov ansatz \cite{DelFerraro2015-92}. While this can be expected to work well for stationary states at high temperatures, such approximations are usually quite severe for short to intermediate times or low temperatures.
Also, for dense networks, where the cavity method is not applicable, approximation schemes like the cluster variational method \cite{Pelizzola2013-86,Vazquez2017-3,Pelizzola2017-7} or perturbative schemes \cite{Roudi2011,Aurell2012-85,BachschmidRomano2016-49} have been developed.

In this paper, we present an efficient novel algorithm for precise solutions of the parallel dynamic cavity equations for generic (locally tree-like) graphs and generic bidirectional dynamics. The main feature is the reduction of the computational complexity from exponential to polynomial in the duration of the dynamical process. The central objects in the dynamic cavity method are conditional probabilities for vertex trajectories of nearest neighbors -- the so-called edge messages. As temporal correlations are decaying in time and/or time difference $|t-t'|$, we exploit that the edge messages can be approximated by matrix products; i.e., there is one matrix for every edge, edge state, and time step, encoding the temporal correlations in the corresponding part of the evolution. It turns out that the dimensions of these matrices do not have to be increased exponentially in time. One can obtain quasi-exact results with relatively small matrix dimensions. Computation costs and accuracy can be tuned by controlling the dimensions through controlled truncations.
The idea of exploiting the decay of temporal correlations to approximate edge messages in matrix product form is in analogy with the use of matrix product states \cite{Accardi1981,Fannes1992-144,Rommer1997,Schollwoeck2011-326,Oseledets2011-33} for the simulation of strongly correlated, mostly one-dimensional, quantum many-body systems. These have been used very successfully in algorithms like the density-matrix renormalization group \cite{White1992-11,White1993-10} to study, for example, quantum ground-state properties, often with machine precision \cite{Schollwoeck2005}.
Besides lifting the restrictions of the aforementioned approaches, the matrix product edge-message (MPEM) algorithm can also outperform Monte Carlo simulations (MC) of the dynamics in important respects. In particular, besides allowing for the simulation of single instances, alternatively, one can work directly in the thermodynamic limit. Perhaps more importantly, it has a favorable error scaling. While statistical errors in MC decay very slowly with the number of samples $N_s$ as $1/\sqrt{N_s}$, MPEM yields also observables with absolutely small expectation values with very good accuracy which is essential for the study of decay processes and temporal correlations. As a first application, we solve the prototypical example for non-equilibrium dynamics on networks -- Glauber dynamics of the kinetic Ising model \cite{Glauber1963-4} -- and study the equilibration of the magnetization as well as temporal correlations.

\Section{The dynamic cavity method}
\begin{figure}[b]
\label{fig:msgSubgraph}
\includegraphics[width=0.45\linewidth]{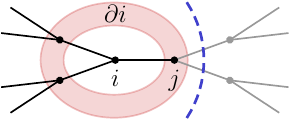}
\caption{Part of a locally tree-like interaction graph with vertex degrees $z=3$.}
\end{figure}
Let $\sigma^t_i$ denote the state of vertex $i$ at time step $t$, and $\vs^t:=(\sigma^t_1,\sigma^t_2,\dots)$ the state of the full system at time $t$. Given the state probabilities $P(\vs^t)$ for time $t$, we evolve to the next time step, $P(\vs^{t+1})=\sum_{\vs^t} W(\vs^{t+1}|\vs^t) P(\vs^t)$, by applying the stochastic transition matrix $W$. As vertex $i$ interacts only with its nearest neighbors $j\in\partial i$, the probability for $\sigma_i^{t+1}$ only depends on the states $\sigma_j^{t}$ of these vertices at the previous time step such that the global transition matrix $W$ is a product of local transition matrices $w_i$,
\begin{equation}\label{eq:transitionMatrix}
	W(\vs^{t+1}|\vs^t) = \prod_{i} w_i(\sigma_i^{t+1}|\vs_{\partial i}^{t}).
\end{equation}
Here $\sum_{\sigma_i} w_i(\sigma_i|\vs'_{\partial i})=1$, and $\vs_{\partial i}^{t}$ is the state of the nearest neighbors of vertex $i$ at time $t$.
In the cavity method \cite{Mezard2001-20,Neri2009-08,Karrer2010-82}, one neglects cycles of the (locally tree-like) graph according to the Bethe-Peierls approximation to reduce this computationally complex evolution to the dynamic cavity equation \cite{Neri2009-08,Karrer2010-82}
\begin{multline}\label{eq:cavityEquation}
	\mu_{i\to j}(\bs_i^{t+1}|\bs_j^{t}) = \hspace{-0.5ex}\sum_{\{\bs_k^t\}_{k\in \partial i\setminus\{j\}}}\hspace{-1ex}  p_i(\sigma_i^0)
	  \Big[\prod_{s=0}^t w_i(\sigma_i^{s+1}|\vs_{\partial i}^s)\Big]\\
	  \times \Big[\prod_{k\in\partial i\setminus\{j\}} \mu_{k\to i}(\bs_k^{t}|\bs_i^{t-1})\Big]
\end{multline}
which only involves the so-called edge messages $\mu$ for the edges of a single vertex $i$. For simplicity, we have assumed that vertices are uncorrelated in the initial state such that 	$P(\vs^0)=\prod_{i}p_i(\sigma_i^0)$.
The edge messages $\mu_{i\to j}(\bs_i^{t}|\bs_j^{t-1})$ in the dynamic cavity equation \eqref{eq:cavityEquation} are conditional probabilities for the trajectories $\bs^t_i:=(\sigma^0_i,\sigma^1_i,\dots,\sigma^t_i)$ and $\bs^{t-1}_j$ on edge $(i,j)$. Specifically, if we consider a tree graph and remove all descendants of vertex $j$ as indicated in Figure~\ref{fig:msgSubgraph} by the dashed line, $\mu_{i\to j}(\bs_i^{t}|\bs_j^{t-1})$ denotes the conditional probability of a trajectory $\bs_i^{t}$ on vertex $i$, given the trajectory $\bs_j^{t-1}$ on vertex $j$. From messages, one can obtain marginal probabilities of site trajectories to evaluate observables of interest. Equation~\eqref{eq:cavityEquation} constructs $\mu_{i\to j}(\bs_i^{t+1}|\bs_j^{t})$ out of the edge messages $\mu_{k\to i}(\bs_k^{t}|\bs_i^{t-1})$ of the previous time step. This is exact for tree graphs and covers locally tree-like graphs in the Bethe-Peierls approximation.
Although we have gained a lot in the sense that the computational complexity is now linear in the system size, it is still exponential in time $t$, if we were to encode the edge messages without any approximation.

\Section{Canonical form of an MPEM}
To circumvent this exponential increase of computation costs, we can exploit the decay of temporal correlations and approximate the exact edge message by a matrix product
\begin{multline}\label{eq:MPEM}
	\mu_{i\to j}(\bs_i^{t}|\bs_j^{t-1}) = A^{(0)}_{i\to j}(\sigma_j^0) \Big[\prod_{s=1}^{t-1} A^{(s)}_{i\to j}(\sigma_i^{s-1}|\sigma_j^s)\Big]\\
	\times A^{(t)}_{i\to j}(\sigma_i^{t-1})A^{(t+1)}_{i\to j}(\sigma_i^{t}).
\end{multline}
The particular choice of assigning vertex variables $\{\sigma^s_i\}$ and $\{\sigma^s_j\}$ to the $M_s\times M_{s+1}$ matrices $A^{(s)}_{i\to j}(\sigma_i^{s-1}|\sigma_j^s)$ occurring in the matrix product \eqref{eq:MPEM} is advantageous for the implementation of the recursion relation \eqref{eq:cavityEquation} for MPEMs, as will become clear in the following. In order for the matrix product to yield a scalar, we set $M_0=M_{t+2}=1$.

\Section{MPEM evolution}
\begin{figure*}[t]
\label{fig:MPEM}
\includegraphics[width=\textwidth]{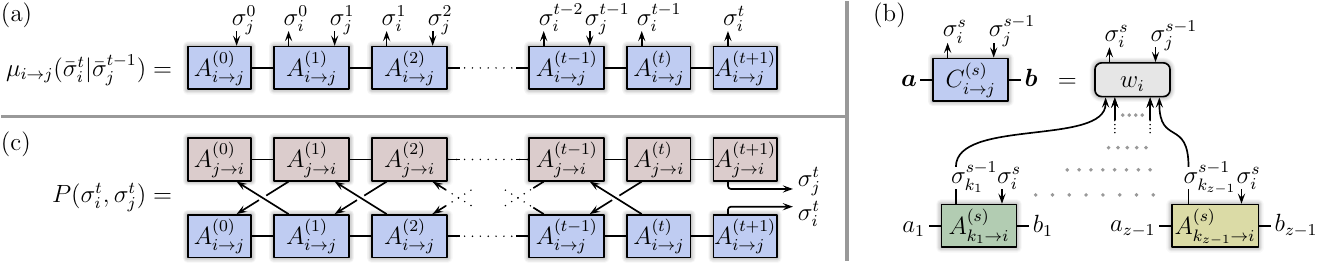}
\caption{(a) Graphical representation of a matrix product edge message in canonical form \eqref{eq:MPEM}. Connecting lines indicate summations over indices. (b) For each time step \eqref{eq:cavityEquation}, tensors of the evolved matrix product $\mu_{i\to j}(\bs_i^{t+1}|\bs_j^{t})$ in Eq.~\eqref{eq:MPEM_C} are built by contracting the local transition matrix $w_i$ with MPEM tensors of messages $\mu_{k\to i}$, incident to vertex $i$, where $k\in\partial i\setminus\{j\}=\{k_1,\dotsc,k_{z-1}\}$, and $\vec{a}:=(a_1,\dotsc,a_{z-1})$. (c) Evaluation of probabilities $P(\sigma_i^t,\sigma_j^t)$ as in Eq.~\eqref{eq:evalP}.}
\end{figure*}
The time evolution starts at $t=0$ with $\mu_{i\to j}(\sigma_i^{0})=p_i(\sigma^0_i)$. Using the dynamic cavity equation \eqref{eq:cavityEquation}, we iteratively build matrix product approximations for edge messages for time $t+1$ from those for time $t$. It is simple to insert the matrix product ansatz \eqref{eq:MPEM} for the edge messages in the dynamic cavity equation, but not trivial to bring the resulting edge message again into the canonical MPEM form as required for the subsequent evolution step. The specific assignment of the vertex variables to matrices in Eq.~\eqref{eq:MPEM} has been chosen such that all contractions (products and sums over vertex variables) occurring in the cavity equation are time-local in the sense that, given MPEMs $\mu_{k\to i}(\bs_k^{t}|\bs_i^{t-1})$ in canonical form for all neighbors $k\in\partial i\setminus\{j\}$, the resulting $\mu_{i\to j}(\bs_k^{t+1}|\bs_i^{t})$ can be written in (non-canonical) matrix product form as
\begin{equation}\label{eq:MPEM_C}
	\mu_{i\to j}(\bs_i^{t+1}|\bs_j^{t}) = C^{(0)}_{i\to j}(\sigma_i^{0}) \Big[\prod_{s=1}^{t+1} C^{(s)}_{i\to j}(\sigma_i^{s}|\sigma_j^{s-1})\Big].
\end{equation}
As depicted in Figure~\ref{fig:MPEM}b, the tensors $C^{(s)}_{i\to j}$ for $1\leq s\leq t$ are obtained by contracting the local transition matrix $w_i(\sigma_i^{s}|\vs_{\partial i}^{s-1})$ with tensors $A^{(s)}_{k\to i}$ from the time-$t$ MPEMs. This contraction entails a sum over the $z-1$ common indices $\sigma_k^{s-1}$, where $z=|\partial i|$ is the vertex degree. Assuming for the simplicity of notation that the matrix dimensions $M_s$ for all time-$t$ MPEMs are identical, the resulting matrices
\begin{multline}
	C^{(s)}_{i\to j}(\sigma_i^{s}|\sigma_j^{s-1})
	= \sum_{\vs_{\partial i\setminus\{j\}}^{s-1}}
	    w_i(\sigma_i^{s}|\vs_{\partial i}^{s-1})\\
	    {\times}
	  \Big[\bigotimes_{k\in\partial i\setminus\{j\}} A^{(s)}_{k\to i}(\sigma_{k}^{s-1}|\sigma_i^s)\Big]
\end{multline}
have left and right indices of dimensions $\bar{M}_s=(M_s)^{z-1}$ and $\bar{M}_{s+1}=(M_{s+1})^{z-1}$, respectively. The contraction for tensor $C^{(t+1)}$ is very similar and
$	C^{(0)}_{i\to j}(\sigma_i^{0})
	= p_i(\sigma_i^0)\allowbreak
	  \big[\bigotimes_{k\in\partial i\setminus\{j\}} A^{(0)}_{k\to i}(\sigma_i^0)\big]$.

\Section{MPEM truncation}
In preparation for the next time step, we need to bring the evolved edge message \eqref{eq:MPEM_C} back into canonical form \eqref{eq:MPEM}. Furthermore, we need to introduce a controlled approximation that reduces the matrix dimensions because they would otherwise grow exponentially in time. Both, a reordering of the vertex variables $\sigma_i^s$ and $\sigma_j^s$ in the matrix product \eqref{eq:MPEM_C} and a controlled truncation of matrix dimensions can be achieved by sweeping through the matrix product and doing certain singular value decompositions (SVD) \cite{Golub1996} of the tensors $C^{(s)}$. The generic idea behind the truncation of a matrix product  $\gamma(\vn):=A_0^{n_0} A_1^{n_1}\dotsb A_t^{n_t}$ with matrix dimensions $\{M_s\}$ is to part the variables $\vn$ into two groups $\vn_L:=(n_0,\dotsc,n_{r-1})$ and $\vn_R:=(n_r,\dotsc,n_{t})$ such that $\Gamma_{\vn_L,\vn_R}:=\gamma(\vn)$ can be interpreted as a matrix. Its singular value decomposition has the form $\Gamma_{\vn_L,\vn_R}=\sum_{k=1}^{M_r} Y_{\vn_L,k}\lambda_k Z_{k,\vn_R}$ with isometric matrices $Y$ and $Z$. Retaining only the $M'_r\leq M_r$ largest singular values $\lambda_k$, we obtain a controlled approximation $\gamma_\trunc(\vn)$ of the original matrix product $\gamma(\vn)$ with 2-norm distance $\norm{\gamma-\gamma_\trunc}^2=\sum_{k>M'_r}\lambda_k^2$ and decreased matrix dimension $M'_r$ at the (temporal) bond $(r-1,r)$.

Following this principle, the truncation of all matrix dimensions of the time-evolved MPEM \eqref{eq:MPEM_C} can be done by sequential SVDs of tensors in the matrix product. In a first sweep, starting with a decomposition of the rightmost tensor $C^{(t+1)}_{i\to j}(\sigma_i^{t+1}|\sigma_j^{t})\stackrel{\SVD}{=:}U^{(t+1)}\Lambda^{(t+1)}\tilde{C}^{(t+1)}_{i\to j}(\sigma_i^{t+1}|\sigma_j^{t})$ and progressing iteratively to the left with 
\begin{equation*}
	C^{(s)}_{i\to j}(\sigma_i^{s}|\sigma_j^{s-1})U^{(s+1)}\Lambda^{(s+1)}\stackrel{\SVD}{=:}U^{(s)}\Lambda^{(s)}\tilde{C}^{(s)}_{i\to j}(\sigma_i^{s}|\sigma_j^{s-1}),
\end{equation*}
the new tensors $\tilde{C}$ are isometries. In a second sweep from left to right, matrix dimensions can be truncated from $\bar{M}_s$ to something smaller in an SVD. In two further sweeps, the indices $\{\sigma_i^s\}$ and $\{\sigma_j^s\}$ of the vertex variables can be rearranged to get back to the canonical form \eqref{eq:MPEM}. After executing these steps for all edge messages, the next evolution step from $t+1$ to $t+2$ can follow. More details of this procedure are described in the appendices.
\begin{figure*}[t]
\label{fig:Glauber}
\includegraphics[width=\textwidth]{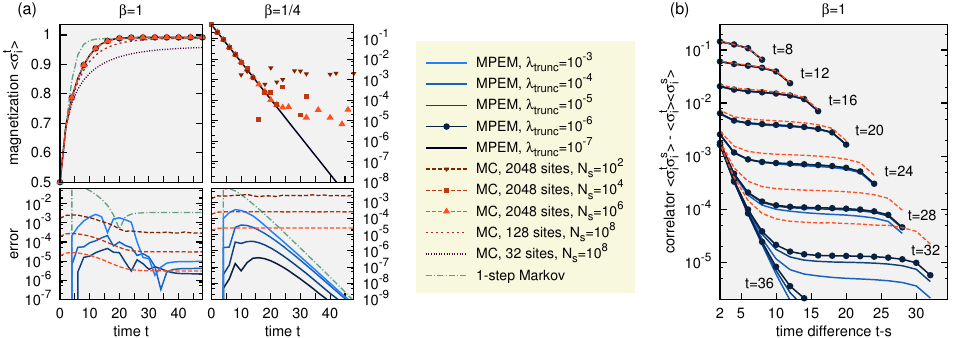}
\caption{(a) Magnetization and (b) connected temporal correlations for Glauber dynamics on $z=3$ random regular graphs of different sizes for MC and in the thermodynamic limit for the MPEM and one-step Markov approaches. Because of odd-even effects in the dynamics, only even time steps are shown. For MC (with $N_s$ samples), the errors of the magnetization [lower panels in (a)] are quantified by the standard deviation of the magnetization, i.e., under ignorance of remaining finite-size effects. For MPEM and one-step Markov, errors are quantified by the deviation from the result of the most accurate (quasi-exact) MPEM simulation (truncation threshold $\lambda_\trunc=10^{-6}$ for $\beta=1$ and $\lambda_\trunc=10^{-7}$ for $\beta=1/4$). In plot (b), the three MPEM curves for $\lambda_\trunc=10^{-4},10^{-5},10^{-6}$ overlap up to time $t=24$.}
\end{figure*}

\Section{Evaluation of observables}
The joint probability of trajectories $\bs_i^t$ and $\bs_j^{t-1}$ for the vertices of an edge $(i,j)$ is given by the product of the two corresponding edge messages. After marginalization, one obtains, for example, the probability for the edge state $(\sigma_i^t,\sigma_j^t)$ at time $t$ as
\begin{equation}\label{eq:evalP}
	P(\sigma_i^t,\sigma_j^t)=\sum_{\bs_i^{t-1},\bs_j^{t-1}} \mu_{i\to j}(\bs_i^t|\bs_j^{t-1}) \mu_{j\to i}(\bs_j^t|\bs_i^{t-1}).
\end{equation}
In the MPEM approach, this can be evaluated efficiently, as indicated in Figure~\ref{fig:MPEM}c, by executing the contractions sequentially from left ($s=0$) to right ($s=t-1$). Similarly, one can for example also compute temporal correlators $\bra\sigma_i^t\sigma_i^{s}\ket$ from probabilities $P(\sigma_i^t,\sigma_i^{s})$.

\Section{Non-equilibrium Glauber-Ising dynamics}
We have used the novel MPEM algorithm to study Glauber dynamics of the kinetic Ising model, introduced in Glauber's seminal paper Ref.~\cite{Glauber1963-4}. Figure~\ref{fig:Glauber} displays the results in comparison to MC simulations and to the one-step Markov approximation \cite{DelFerraro2015-92}. Specifically, we have Ising spins interacting ferromagnetically on $z=3$ random regular graphs, with local transition matrices $w_i(\sigma_i^{t+1}|\vs_{\partial i}^{t})=\exp(\beta\sum_{j\in\partial i} \sigma_i^{t+1}\sigma_j^{t})/Z$. In the initial state, all spins have magnetization $\bra\sigma^0_i\ket=1/2$, i.e., $p_i(\ua)=3/4$. 
Besides being applicable for single instances of finite graphs, the MPEM approach gives also direct access to the thermodynamic limit. For disordered systems, this can be done in a population dynamics scheme. The homogeneous case, considered here, is particularly simple as all edges of the graph are equivalent in the thermodynamic limit. Hence, one can work with a single MPEM.

Figure~\ref{fig:Glauber}a shows the equilibration of the magnetization. In the ferromagnetic phase ($\beta=1$), it approaches a finite equilibrium value, whereas it decays exponentially to zero in the paramagnetic phase ($\beta=1/4$). As shown for $\beta=1$, MC simulations contain finite-size effects which become small for the system with $2048$ sites. MC errors decrease slowly when increasing the number of samples $N_s$ as $1/\sqrt{N_s}$. This is problematic for observables with small absolute values where cancellation effects make it difficult to obtain a precise estimate. This is, e.g., apparent in the magnetization decay for $\beta=1/4$ which, in contrast, is very accurately captured with MPEM. In these simulations, we control the MPEM accuracy by keeping only singular values $\lambda_k$ above a threshold, specified by $\lambda_k/(\sum_{k'}\lambda_{k'}^2)^{1/2} > \lambda_\trunc$. Decreasing $\lambda_\trunc$, increases accuracy and computation costs. The one-step Markov approximation \cite{DelFerraro2015-92} is not suited to handle temporal correlations. At long times it performs well for $\beta=1/4$ and fairly good for $\beta=1$, but deviates rather strongly at earlier times.

Figure~\ref{fig:Glauber}b shows the connected temporal correlation function $\bra\sigma_i^t\sigma_i^{s}\ket-\bra\sigma_i^t\ket\bra\sigma_i^{s}\ket$ for the ferromagnetic regime $\beta=1$ as a function of $t-s$ for several times $t$. After an initial exponential decay in $t-s$ to a value that decreases exponentially with $t$, the correlator continues to drop but now much more slowly, becoming almost constant. Its decay behavior can be difficult to impossible to capture with MC. In the example, MC deviations are often orders of magnitude above those of the numerically cheaper MPEM simulations. MC data for $t>32$ have been suppressed due to very big errors.

This decay of temporal correlations is also reflected in the matrix dimensions $\{M_s\}$ in MPEMs $\mu_{i\to j}(\bs_i^{t}|\bs_j^{t-1})$ of predefined approximation accuracy. We observe that $M:=\max_s M_s$ increases rapidly at small times $t$ but then converges to a value that depends on the system parameters and the truncation threshold $\lambda_\trunc$. As every iteration $t\to t+1$ requires a few sweeps through matrix products of length $t$, this implies that computation costs are $\mathcal{O}(t^2)$, i.e., quadratic instead of exponential in $t$.

\Section{Discussion}
The novel MPEM algorithm, based on matrix product approximations of edge messages allows for an efficient and accurate solution of the dynamic cavity equations. Besides lifting restrictions of earlier approaches for the simulation of stochastic non-equilibrium dynamics in networks, mentioned in the introduction, it gives direct access to the thermodynamic limit, and its error scaling is favorable to that of MC simulations. It allowed us to obtain quasi-exact solutions of the cavity equations for Glauber-Ising dynamics. We think that this new approach is a very valuable tool, particularly as it yields temporal correlations and other decaying observables with unprecedented accuracy as demonstrated in the example. It hence gives access to low-probability events. This opens a new door for the study of diverse dynamic processes and inference or dynamic optimization problems for physical, technological, biological, and social networks.

We thank G.~Del Ferraro for providing data to benchmark our one-step Markov simulation for Figure~\ref{fig:Glauber} and acknowledge support by the Marie Curie Training Network NETADIS (FP7, grant 290038). 

\newpage
\appendix

\section{Truncating matrix products}\label{appx:truncate}\vspace{-0.3em}
Let us explain the notion of truncations at the example of a matrix product
\begin{equation}\label{eq:MP}
	\gamma(\vn):=A_0^{n_0} A_1^{n_1}\dotsb A_t^{n_t},
\end{equation}
where $A_s^{n_s}$ is an $M_s\times M_{s+1}$ matrix and $M_0=M_{t+1}=1$. Our goal is to reduce in a controlled way, e.g., the left matrix dimension $M_{r}$ of $A_r^{n_r}$. First, let us part the variables $\vn$ into two groups $\vn_L:=(n_0,\dotsc,n_{r-1})$ and $\vn_R:=(n_r,\dotsc,n_{t})$. For the truncation, we suggest to employ a singular value decomposition (SVD) \cite{Golub1996} of the matrix product such that 
\begin{equation}\label{eq:SVD}
	\gamma(\vn)=: \Gamma_{\vn_L,\vn_R}\stackrel{\text{SVD}}{=}\,\sum_{k=1}^{M_r} Y_{\vn_L,k}\lambda_k Z_{k,\vn_R}
\end{equation}
$Y$ and $Z$ are isometric matrices, i.e., they obey
\begin{equation}
	Y^\dag Y=\id \quad\text{and}\quad Z Z^\dag=\id.
\end{equation}
Now, truncating some of the singular values $\lambda_1\geq \lambda_2\geq\dots \geq \lambda_{M_r}\geq 0$, such that only the $M_r'$ largest are retained, we obtain the controlled approximation
\begin{multline}\label{eq:gammaTrunc}
	\gamma_\trunc(\vn):= \sum_{k\leq M_r'} Y_{\vn_L,k}\lambda_k Z_{k,\vn_R}\\
	\text{with error}\quad
	\norm{\gamma-\gamma_\trunc}^2=\sum_{k> M_r'}\lambda_k^2.
\end{multline}
Note that this truncation scheme yields the minimum possible norm loss $\norm{\Delta\gamma}\equiv\left(\sum_\vn \Delta\gamma^2(\vn)\right)^{1/2}$ for the given new matrix dimension $M_r'$.

While it is very desirable to discard unimportant information and control the growth of computation cost through such truncations, the SVD \eqref{eq:SVD} appears to be an insurmountable task. Assuming that each variable $n_s$ can take $d$ different values and that $2r\leq t+1$, the cost for the SVD would scale exponentially in time like $d^{t+r+1}$. This is because the SVD of an $M\times N$ matrix with $M\leq N$ has a computation cost $\mathcal{O}(M^2N)$ \cite{Golub1996}. However, the beauty of matrix products is that such an SVD can in fact be done sequentially with linear costs of order $tdM^3$ as follows. Here, $M:=\max_s M_s$ is the maximum matrix dimension in Eq.~\eqref{eq:MP}.

First, we do an exact transformation of the matrix product \eqref{eq:MP} to bring it to the orthonormalized form
\begin{equation}\label{eq:MP-ON}
	\gamma(\vec{n})=Y_0^{n_0}\dotsb Y_{r-1}^{n_{r-1}} \tilde{A}_r^{n_r} Z_{r+1}^{n_{r+1}}\dotsb Z_t^{n_t},
\end{equation}
where tensors $Y_s$ and $Z_s$ obey the left and right orthonormality constraints
\begin{equation}\label{eq:ON}
	\sum_n (Y_s^n)^\dag Y_s^n = \id\quad\text{and}\quad
	\sum_n Z_s^n (Z_s^n)^\dag = \id,
\end{equation}
respectively. This is achieved through a sequence of SVDs. It starts with the SVD $A^{n_0}_0=:Y^{n_0}_0\Lambda_0 V_0$, where $\Lambda_0$ is a diagonal matrix containing the singular values, $V_0$ is isometric according to $V_0V_0^\dag=\id$, and $Y_0$ obeys Eq.~\eqref{eq:ON}. The sweep continues with the SVD $\Lambda_0 V_0 A^{n_1}_1=:Y^{n_1}_1\Lambda_1 V_1$ and so on until the computation of $Y_{r-1}$. Analogously, we do a second sequence of SVDs starting from the right
with $A^{n_t}_t=:U_t\Lambda_t Z_t^{n_t}$, where $Z_t^{n_t}$ obeys Eq.~\eqref{eq:ON}, and continue with $A^{n_{t-1}}_{t-1}U_t\Lambda_t=:U_{t-1}\Lambda_{t-1} Z_{t-1}^{n_{t-1}}$ and so on until $Z_{r+1}$ has been computed.
Finally, we define the central tensor as $\tilde{A}_r^{n_r}:=\Lambda_{r-1} V_{r-1} A_r^{n_r}U_{r+1}\Lambda_{r+1}$ and have thus determined all matrices in Eq.~\eqref{eq:MP-ON}. After this preparation, we can do the actual truncation, based on the SVD $\tilde{A}_r^{n_r}=:U_r\Lambda Z^{n_r}_r$ with the same singular values $\lambda_1\geq \dots\geq \lambda_{M_r}$ as in Eq.~\eqref{eq:SVD}. With the $M_r\times M_r'$ matrix $[\Lambda_\trunc]_{kk'}:=\delta_{kk'}\lambda_k$, the truncated matrix product \eqref{eq:gammaTrunc} takes the form
\begin{equation*}
	\gamma_\trunc(\vn)= Y_0^{n_0} \dotsb Y^{n_{r-2}}_{r-2} (Y^{n_{r-1}}_{r-1} U_r \Lambda_\trunc)Z_r^{n_r}\dotsb Z_t^{n_t}.
\end{equation*}

\section{Processing evolved MPEMs}\label{appx:processingMPEM}\vspace{-0.3em}
In the evolution step described in the main text, matrix dimensions are increased to $\bar{M}_s$ and the evolved edge message \eqref{eq:MPEM_C}
is in a non-canonical form. Here, we discuss how to apply the truncation as described in  Appendix~\ref{appx:truncate} to compress the evolved MPEM and bring it back to canonical form \eqref{eq:MPEM}.

In a first sweep from right ($s=t+1$) to left ($s=0$), using SVDs, we can sequentially impose the right orthonormality constraints [see Eq.~\eqref{eq:ON}] on the $C$-tensors. In a subsequent sweep from left to right, again based on SVDs, at each step, the MPEM is in orthonormalized form [see Eq.~\eqref{eq:MP-ON}] and
we can now truncate the tensors to decrease bond dimensions from $\bar{M}_s$ to something smaller. According to the triangle inequality, the norm distance of the original edge message $\mu_{i\to j}(\bs_i^{t+1}|\bs_j^{t})$ and the resulting truncated MPEM are bounded from above by the sum of errors \eqref{eq:gammaTrunc} of the individual truncations.

What remains is to reorder the indices $\{\sigma_i^s\}$ and $\{\sigma_j^s\}$ of the vertex variables. In a sweep from right to left, we go from the non-canonical variable assignment $(\sigma_i^{0})(\sigma_i^{1}|\sigma_j^{0})\dots(\sigma_i^{t+1}|\sigma_j^{t})$ in the truncated and orthonormalized version $\tilde{C}^{(0)}_{i\to j}(\sigma_i^{0}) \prod_{s=1}^{t+1} \tilde{C}^{(s)}_{i\to j}(\sigma_i^{s}|\sigma_j^{s-1})$ of the MPEM \eqref{eq:MPEM_C} to the assignment $(\sigma_i^{0}\sigma_j^{0})\dots(\sigma_i^{t}|\sigma_j^{t})(\sigma_i^{t+1})$, yielding the matrix product
\begin{equation*}
	\mu_{i\to j}(\bs_i^{t+1}|\bs_j^{t}) \stackrel{\trunc}{\approx} \Big[\prod_{s=0}^{t} D^{(s)}_{i\to j}(\sigma_i^{s}|\sigma_j^{s})\Big] D^{(t+1)}_{i\to j}(\sigma_i^{t+1}).
\end{equation*}
At the right boundary, we start with an SVD and controlled truncation $\tilde{C}^{(t+1)}_{i\to j}(\sigma_i^{t+1}|\sigma_j^{t})\allowbreak\approx: U^{(t+1)}(\sigma_j^{t})\allowbreak {\times}\Lambda_\trunc^{(t+1)}\allowbreak D^{(t+1)}_{i\to j}(\sigma_i^{t+1})$, and continue with $\tilde{C}^{(t)}_{i\to j}(\sigma_i^{t}|\sigma_j^{t-1})\allowbreak {\times}U^{(t+1)}(\sigma_j^{t})\allowbreak\Lambda_\trunc^{(t+1)}\allowbreak\approx: U^{(t)}(\sigma_j^{t-1})\allowbreak\Lambda_\trunc^{(t)}\allowbreak D^{(t)}_{i\to j}(\sigma_i^{t}|\sigma_j^{t})$ and so on until ending at $s=0$. In an analogous final sweep from left to right, we change to the canonical variable assignment $(\sigma_j^{0})(\sigma_i^{0}|\sigma_j^{1})\dots(\sigma_i^{t-1}|\sigma_j^{t})(\sigma_i^{t})(\sigma_i^{t+1})$ as in Eq.~\eqref{eq:MPEM}. After executing these steps for all edge messages, the next evolution step from $t+1$ to $t+2$ can follow.

\end{document}